\documentclass[aps,pra,preprint,groupedaddress]{revtex4}
\usepackage{amssymb, amsmath}
\usepackage{graphicx}
\begin{document}

%\preprint{}

%\title{M\'{e}thodologie pour d\'{e}doubler les Bosons identiques dans un pi\`{e}ge avec deux puits}
%\title{General Computational Theory of a splitting Bose-Einstein Condensate I: Theory and Statics}
\title{Computational Theory of a splitting BEC using a Generalized Wannier basis I: Theory and Statics}



\author{Douglas K. Faust}
\email[]{dfaust@phys.washington.edu}
\affiliation{Department of Physics, University of Washington, Seattle, Washington 98195-1560, USA}
\author{William P. Reinhardt}
\email[]{rein@chem.washington.edu}
\affiliation{Department of Physics, University of Washington, Seattle, Washington 98195-1560, USA}
\affiliation{Department of Chemistry, University of Washington, Seattle, Washington 98195-1700, USA}


\date{\today}

\begin{abstract}

We investigate the behavior of a Bose-Einstein Condensate (BEC) under the influence of a central barrier as the particle number trends towards the thermodynamic limit.  In 
order to perform these studies, we present a novel method which is tractable in the large-$N$ limit.  This method employs what may be considered to be 
a generalized Wannier basis, which successfully incorporates features of previous theoretical and computational assays to the splitting problem, including
mean field effects, and has access to the dimensionality, trap parameters, and particle numbers relevant to recent experiments.  At any barrier height we are able to discern between 
a two-mode state and a state which is described sufficiently by mean field theory and, further, give a criterion and technique for matching the two-mode theory to the 
zero-barrier state.
We compare the basis used in this model to the de-localized basis functions underlying alternate models used in recent theoretical work on the double-well
splitting problem and show that only the generalized Wannier basis displays the level crossing and emergence of two complex order parameters with overall
$U(1) \oplus U(1)$ symmetry as expected from a large-$N$ analogue of the Superfluid to Mott insulator transition.
Using this model, we identify a universal structure, independent of $N$, in this phase transition.   We also present an analytic and model-independent description of this 
universal structure and discuss its consequences for realizing true two-mode physics with a BEC which trends towards the thermodynamic limit.  

\end{abstract}

%\keywords{}

\maketitle

\section{Introduction}\label{Introduction}

The realization of Bose Einstein condensation in a dilute gas of atoms \cite{JILABEC} and the verification that the condensate order parameter both exists, 
and is characterized by the existence of a well-defined global phase, as witnessed by the observations of solitons \cite{soliton}, vortices \cite{Vortex1}\cite{Vortex2} 
and laser-like interference \cite{KetterleBEC} has stimulated a great deal of research in recent years.
Specifically, a BEC coherently split in a double-well potential holds promise as a basic tool to study symmetry breaking, 
decoherence and phase diffusion properties of quantum systems as well as promise for use as an interferometric tool with an efficiency 
below the shot-noise limit \cite{heisenberglimit}.  As such, these systems have attracted a lot of attention, with Saba et al. 
creating the first such BEC interferometer in 2004 \cite{MITinterf}.  These experiments have been refined with atom-chip technology to the point where
a deterministic precession of relative phase \cite{Schmied} attributed to differences in mean particle number in each well has been observed as well as a loss of a 
deterministic interference pattern attributed to phase diffusion \cite{MITsqueeze}.
At smaller particle number Oberthaler et al. obtained fine enough control over a 
central barrier to reach the Josephson regime in 2006 \cite{BJJ}.

One basic scientific question associated with the ``splitting'' of a BEC,
first raised in \cite{MITinterf} is whether, or under what conditions, the two moieties generated after splitting have the same phase and how long such 
phase-coherence lasts. 
Since a BEC is a mesoscopic system it is not \textit{a priori} clear whether a classical or a quantum mechanical description of 
the order parameter is appropriate.  In the former case, two independent condensates split from a common progenitor would share a phase until interaction 
with the environment destroyed phase coherence similar to the way that the pieces of a cleaved crystal would retain a common orientation.  
In the latter case, a quantum description of the splitting process suggests, since relative number and relative phase are conjugate variables, phase coherence 
will be destroyed for independent condensates.  
Lattice experiments operating at small particle per lattice site exhibit a loss of phase coherence during the so-called
Superfluid to Mott insulator transition, but show a restoration of phase-coherence faster than so-called ``phase-incoherent'' states once the lattice depth is 
decreased \cite{SFMI}.
It is not clear, without a theoretical tool to accompany these experiments, what is responsible for these two phenomena, where both the Mott Insulator
state and the phase-coherent state are described as ``fragmented,'' or pure Fock
states with perfectly defined particle number.  Such a theoretical tool should also be able to investigate what barrier heights and ramping times are needed to 
engineer ``squeezed'' and other exotic states.
We introduce such a theoretical method capable of accessing physical data inaccessible to \textit{in situ} imaging
in order to compliment the set of BEC splitting experiments currently being performed in hopes of answering these fundamental and technical questions. 

Theoretical descriptions of certain aspects of degenerate bosonic systems in one or more modes have been contributed from a variety of disciplines within physics 
over the last decade.  
The description of a BEC in terms of an order parameter with definite phase as an example of spontaneous symmetry-breaking is inherited from the 
condensed matter literature, as is the use of model many-body wavefunctions and Fock-space expansion coefficients to describe 
particle distributions in between multiple wells, aka the Bose-Hubbard model. 
When describing a two-mode or splitting BEC, Bose-Hubbard type models use tunneling rates and 
site energies, which are typically input as parameters or approximate wavefunctions \cite{SandS} but may actually have a complicated dependence on underlying 
many-body wavefunctions and trap 
geometry as well as having dynamics of their own during the splitting process.  
More realistically, in the context of a highly-restricted ``quantum optics'' type quasi-Gaussian ansatz, Zoller et al. \cite{zoller} investigated 
general properties of the dynamical splitting of a BEC.  Perhaps most importantly, they noted that because there are two different types of dynamical
variables of interest in the system: the spatial variables which govern how the density of the atomic cloud(s) evolve and the Fock 
space variables which govern whether particles are localize into a single well, there are two different types of adiabaticity, each 
associated with a characteristic timescale.  We further note that since \textit{in situ} imaging of a BEC only probes the atomic density of the cloud, the 
distribution of the Fock space variables of the system must therefore be inferred from the density, for instance by the loss and revival of an interference 
pattern.  The detailed analysis of one number-squeezing lattice experiment \cite{kasevich}, however, has shown that mean-field effects may mimic the 
experimental signature of number-squeezing \cite{McKagan} indicating that 
\textit{in situ} imaging should not be used as a stand-alone diagnostic of the Fock-space distribution of a split or squeezed BEC.
However, an \textit{ab initio} computational method allows this variable, the primary component in engineering quantum degenerate states beyond mean-field theory, to be 
investigated directly, by construction.

Recently, more exact, ``multi-configurational'' schemes \cite{djmTD}\cite{djmTI}\cite{CedTD}\cite{CedTI}\cite{Band}, derived from methods familiar to quantum chemists and 
many-body nuclear theorists have been developed which self-consistently include both mean-field and particle number dynamics for a set of indistinguishable 
bosons with access to more than one mode.  In spite of all of these efforts, a single theory which synthesizes the relevant aspects of these approaches 
and calculates useful quantities like phase diffusion rates and the effect of barrier raising times on the subsequent interference patterns has not yet been presented.

There are two essential difficulties in correctly simulating the splitting of an initially-coherent BEC into two distinct moieties.  
First, is the fact that as the barrier raises, the system acquires an additional degree of freedom which was absent at $t=0$.  Namely, at some point in the 
splitting, a breadth of Fock space expansion coefficients corresponding to the possible distributions of particles in either well, is needed.  Conversely, 
mean field theory, as described by the 
Gross-Pitaevskii equation \cite{Gross}\cite{Pitaevskii} - which is the correct theory at low barrier heights - assumes that all particles are
contained in a single quantum state.  A successful theory, therefore must correctly discriminate in between a one-mode (GP like) and two-mode states in the course of a 
time-dependent
and potentially non-adiabatic process.  The method described in this paper applies the Penrose-Onsager criterion to a form of the reduced density operator 
following \cite{zoller} in order to do this.
The second difficulty is that, if one takes a condensed matter theorist's viewpoint, the complete fragmentation of a BEC into two independent BECs should also be an 
example of symmetry breaking, in which a system described by a single order parameter with a global $U(1)$ symmetry, becomes a system with $U(1) \oplus U(1)$ symmetry.  
That is, each of the two wells should contain an independent condensate such that an arbitrary phase shift on one should not affect the total system energy.
In order to motivate our choice of spatial basis functions, we show that this latter requirement is not satisfied by a na\"{i}ve implementation of even the sophisticated 
self-consistent methods such as in \cite{ced_well}. 
In response to these difficulties, we present a basis and method which smoothly transitions from a single totally-occupied zero-temperature ground state to a fragmented state 
in such a way that the initial density as well as the final densities and Fock-space distributions have physical meaning.

In Section \ref{Theory}, we develop such a theory which is correct in both the low and high-barrier limits.  First, we develop equations of motion for two spatial 
mode functions and the possible partitions of particles in between them which are tractable and separable in the limit that $N \gg 1$.  We give a criterion for when 
the system is described by a single totally-occupied state despite using an explicitly two-mode basis and give a further means of checking/correcting all necessary 
physical data of the system in that case.  In Section \ref{Implementation} we briefly describe the numerical implementation of the method described in \ref{Theory}.

% REWRITE
%In Section \ref{Basis} we show results from a set of computations which verify that the basis we have employed, the generalized Wannier basis, correctly reproduces the 
%Superfluid to Mott insulator transition (SF-MI).  Further, we model the same system with a de-localized basis and show that no such curve crossing is present, and point 
%out a second pathology associated with the de-localized basis present when such a basis is used to create approximate left- and right-localized basis functions. 
%In order to understand these how these specific computations will generalize to other systems present estimates of how the band-gap of a de-localized basis scales with
%particle number, trap frequency and interaction strength.  We show that this is strictly positive for repulsive condensates and, as such, the SF-MI transition cannot be
%simulated in models employing such a de-localized basis.
%On the other hand, an analogous calculation in the generalized Wannier basis model shows that, at high barriers, a Mott Insulator phase is always the ground state 
%for repulsive interactions.  Therefore, the curve crossing, and therefore SF-MI transition, is a universal feature of this model.
In Section \ref{Basis} we show results from a set of calculations which verify that the basis we have employed, the generalized Wannier basis, correctly reproduces the Superfluid to Mott insulator transition (SF-MI).  In order to do this, we define what it means within this theory to be in a Superfluid state and within a Mott insulator state, then we show that a curve crossing exists in which the energy of the Mott insulator state eventually falls below that of the Superfluid state as a function of barrier height.
Conversely, we show that in the same physical system treated with a de-localized basis, the Superfluid state always lies energetically below what is defined as the ``fragmented'' state.  Therefore, no such curve crossing exists within the analogous ``single-'' and ``double-macroscopically'' occupied \textit{gerade}/\textit{ungerade} states.
Finally, in order to understand how these specific computations will generalize to other systems, we present estimates of how the band-gap of a de-localized basis scales with particle number, trap frequency and interaction strength.  We show that this is strictly positive for repulsive condensates and, as such, the SF-MI transition cannot be simulated in models employing such a de-localized basis.
On the other hand, an analogous calculation for the generalized Wannier basis model shows that, at high barriers, a Mott insulator phase is always the ground state for repulsive interactions.  Therefore, the curve crossing, and hence the SF-MI transition, is a universal feature of this model.

Our initial physical results from this theory, described in \ref{Results}, follow from a set of numerical and analytical investigations of the splitting process as the number 
of trapped particles trends towards the large-$N$ limit.  We are able to delineate the regime in which a two-mode model is appropriate and this analysis indicates that there 
is a very narrow region, characterized by a universal mathematical structure, in which two-mode models are applicable to splitting process. 

In Appendix A a method to generate an effective $1D$ equation which self-consistently incorporates trap and mean-field data from the transverse directions when there 
is only one principal (splitting) axis of interest is described.

\section{Theory}\label{Theory}

Here we develop the basic dynamical laws which we use to describe a set of $N$ identical bosons in an external trapping potential $V_{ext}(r)$ and give the means to 
interpret the state vector.  

\subsection{State Vector and Equations of Motion}
Starting from the second-quantized Hamiltonian in the contact-approximation for quasi-1D (see \ref{effective1d} for a discussion of how we self-consistently include 
transverse trap data when only one splitting axis, $r$, is of interest):

\begin{equation}
\hat{H} = \displaystyle\int\!dr[\hat{\Psi}^{\dagger}(r)(\hat{T}+\hat{V}_{ext}(r))\hat{\Psi}(r)+\frac{g}{2}\hat{\Psi}^{\dagger}(r)\hat{\Psi}^{\dagger}(r)\hat{\Psi}(r)\hat{\Psi}(r)],
\end{equation}

where $g$ gives the interaction strength through the s-wave scattering length, $a_{s}$, as $g = \frac{4\pi a_{s}\hbar^{2}}{M}$ as is appropriate for a dilute
low-temperature gas and the $\hat{\Psi}$ are the bosonic field operators satisfying the usual commutation relations 
$[\hat{\Psi}(r),\hat{\Psi}(r')] = [\hat{\Psi}^{\dagger}(r),\hat{\Psi}^{\dagger}(r')] = 0$, and $[\hat{\Psi}(r),\hat{\Psi}^{\dagger}(r')] = \delta(|r-r'|)$.
Because we are including the possibility that the state is a correlated two-mode or fragmented state, we must use two-mode field operators 
$\hat{\Psi}(r) = \hat{a}_{1}\phi_{1}(r) + \hat{a}_{2}\phi_{2}(r)$.

To describe all possibilities for the occupation of these modes, the full state vector is
\begin{equation}\label{CIexpansion}
|\Phi\rangle = 
\sum_{\alpha=0}^{N}C_{\alpha}\frac{(\hat{a}^{\dagger}_{1})^{\alpha}}{\sqrt{\alpha !}} \frac{(\hat{a}^{\dagger}_{2})^{N-\alpha}}{\sqrt{(N-\alpha)!}} |vacuum\rangle
\equiv \sum_{\alpha=0}^{N}C_{\alpha}|\alpha\rangle.
\end{equation}

At this point the standard two-mode operator algebra for bosons \cite{SandS} gives
\begin{equation}
E = \langle \Phi |\hat{H}|\Phi \rangle = \sum_{j,k=1}^{2} \rho_{jk}\epsilon_{jk} + \frac{g}{2}\sum_{j,k,l,m=1}^{2}\rho_{jklm}\Gamma_{jklm},
\end{equation}
with the auxiliary definitions $\rho_{jk} \equiv \langle \Phi |\hat{a}^{\dagger}_{j}\hat{a}_{k}| \Phi \rangle$, 
$\rho_{jklm} \equiv \langle \Phi |\hat{a}^{\dagger}_{j}\hat{a}^{\dagger}_{k}\hat{a}_{l}\hat{a}_{m}| \Phi \rangle$, \\
$\epsilon_{jk} \equiv \displaystyle\int\! dr \phi^{*}_{j}(\frac{-\hbar^2}{2m}\nabla^{2}+V_{ext})\phi_{k}$, and
$\Gamma_{jklm} \equiv \displaystyle\int\! dr \phi^{*}_{j}(r)\phi^{*}_{k}(r)\phi_{l}(r)\phi_{m}(r)$.

We now construct the Action in order to generate equations of motion for all of the quantities declared to be dynamical variables
\begin{equation}
S = \displaystyle\int\!dt \lbrace \langle \Phi | \hat{H} - i\hbar\frac{\partial}{\partial t} - \sum_{j,k=1}^{2}(\mu_{jk}\int\!dr\phi_{j}^{\ast}\phi_{k}-\delta_{jk}) | \Phi \rangle \rbrace.
\end{equation}
where the Lagrange multipliers $\mu_{jk}$ are introduced to enforce the constraint $\displaystyle\int\!dr\phi_{j}^{\ast}\phi_{k} = \delta_{jk}$

Using the above
\begin{equation}
S = \displaystyle\int\!dt \lbrace E  - i\hbar\langle \Phi | \frac{\partial}{\partial t} | \Phi \rangle - \sum_{j,k=1}^{2}(\mu_{jk}\int\!dr\phi_{j}^{\ast}\phi_{k}-\delta_{jk}) \rbrace,
\end{equation}

we now declare the set of Fock-space coefficients $\lbrace C_{\alpha}\rbrace$ and the spatial functions $\lbrace \phi_{j}(r) \rbrace$ to be dynamical variables.  
Employing the Dirac-Frenkel variational principle, as is done for the case of arbitrary modes \cite{CedTD}, the equations
of motion for the Fock-space coefficients are given by the condition $\frac{\partial S}{\partial C_{\gamma}^{\ast}} = 0$ gives
\begin{equation}
\frac{\partial S}{\partial C_{\gamma}^{\ast}} = 0 \Leftrightarrow \frac{\partial E}{\partial C_{\gamma}^{\ast}} 
- i\hbar \sum_{\alpha,\beta=0}^{N}\frac{\partial C_{\alpha}^{\ast}}{\partial C_{\gamma}^{\ast}}\dot{C}_{\beta}\langle \alpha | \beta \rangle = 0,
\end{equation}

\begin{equation}
i\hbar\dot{C}_{\gamma} = \sum_{j,k=1}^{2} \frac{\partial \rho_{jk}}{\partial C_{\gamma}^{\ast}}\epsilon_{jk} + 
\frac{g}{2}\sum_{j,k,l,m=1}^{2}\frac{\partial \rho_{jklm}}{\partial C_{\gamma}^{\ast}}\Gamma_{jklm}
\end{equation}
or, in a more compact notation
\begin{equation}\label{CIeom}
i\hbar\dot{C}_{\gamma} = \sum_{\beta = 0}^{N}\cal{H}_{\gamma \beta}C_{\beta}
\end{equation}
where 
\begin{equation}\label{FockHam}
{\cal H}_{\gamma \beta} \equiv \langle \gamma | \left\{ \sum_{j,k=1}^{2} \epsilon_{jk}(\hat{a}_{j}^{\dagger}\hat{a}_{k}) + 
\frac{g}{2}\sum_{j,k,l,m=1}^{2}\Gamma_{jklm}(\hat{a}_{j}^{\dagger}\hat{a}_{k}^{\dagger}\hat{a}_{l}\hat{a}_{m}) \right\} | \beta \rangle.
\end{equation}

Similarly, the equations of motion for the two mode functions are given by the condition $\frac{\partial S}{\partial \phi_{q}^{\ast}(r')} = 0$.  Now, using the fact that
$\frac{\partial \phi_{j}^{\ast}(r)}{\partial \phi_{q}^{\ast}(r')} = \delta_{jq}\delta(|r-r'|)$, this condition yields:

\begin{eqnarray}
i\hbar\sum_{j,k=1}^{2}\rho_{jk} \delta_{jq}\dot{\phi}_{k} = & \nonumber \\ 
\sum_{j,k=1}^{2}\rho_{jk}\delta_{jq}(\hat{T}+\hat{V}_{ext})\phi_{k} 
+ \frac{g}{2}& \displaystyle\sum_{j,k,l,m=1}^{2}\rho_{jklm}(\delta_{jq}\phi_{k}^{\ast}\phi_{l}\phi_{m} + \phi_{j}^{\ast}\delta_{kq}\phi_{l}\phi_{m}) 
 - \displaystyle\sum_{j,k=1}^{2}(\mu_{jk}\delta_{jq}\phi_{k})
\end{eqnarray}

\begin{equation}\label{preFocklobeEOM}
i\hbar\sum_{k=1}^{2}\rho_{qk} \dot{\phi}_{k} = \sum_{k=1}^{2}\rho_{qk}(\hat{T}+\hat{V}_{ext})\phi_{k} + 
g \sum_{k,l,m=1}^{2}\rho_{qklm}(\phi_{k}^{\ast}\phi_{l}\phi_{m}) - \sum_{k=1}^{2}(\mu_{qk}\phi_{k}).
\end{equation}

These variational calculations have been performed elsewhere in the context of identical and 
distinguishable bosons \cite{CedTD}\cite{HDMreview} up to this point.  Typically, the derivation continues by multiplying either side of the above 
equation by $(\rho_{qk})^{-1}$ in order to decouple the two time derivatives, however, this produces singular equations of motion when 
$\rho_{qk}$ has a zero eigenvalue (i.e. when only one state is occupied).  Since this is precisely the initial state we wish to consider 
in the case of a single well deformed into two wells, another procedure is employed.  We will observe and comment on some ramifications of using schemes 
which become singular in the case of only one occupied mode in Section \ref{Basis}.

In deciding how to proceed we need to know what approximations to the state vector are relevant in the large $N$ limit. 
We do not parametrize, truncate or approximate the Fock-space variables keeping the entire expansion (\ref{CIexpansion}) and the exact equation of motion (\ref{CIeom}), 
since the degree of coherence and fundamental interpretation of the system will depend vitally on both the breadth and relative phases of the distribution of the 
variables $\{ C_{\alpha} \}$.

In order to decouple the equations of motion for the $\{ \phi_{j} \}$, we note that the approximate form of the distribution of the $ \{ C_{\alpha} \}$ is binomial 
for the case of any ground state configuration of the symmetric double-well if the mode corresponding to $\phi_{1}$ ($\phi_{2}$) 
is approximately left (right) localized.  

More strictly, in the case of repulsive interactions ($g > 0$), the following bound holds:
\begin{equation}
|C_{\alpha}| < \frac{1}{4}^{N/2}\sqrt{\frac{N!}{(\alpha !)(N-\alpha)!}}.
\end{equation}
Therefore, the standard deviation of the distribution of the $ \{ C_{\alpha} \}$ should scale as $\sim \frac{1}{\sqrt{N}}$ and the numerical prefactors on 
(\ref{preFocklobeEOM}) which derive from the quantities $\rho_{jk}$, $\rho_{jklm}$ will be well characterized by their Fock-state values when $N \gg 1$.  Consequently,
in the thermodynamic limit, the equations of motion for the lobes are given, to leading order in $1/N$, by the ``diagonal'' contributions: $\rho_{jj}$, $\rho_{jkjk}$, $\rho_{jkkj}$.

Using this large-$N$ approximation, (\ref{preFocklobeEOM}) becomes, explicitly writing out the components:
\begin{eqnarray}\label{finalEOM}
i\hbar\left( \begin{array}{cc} \rho_{11} & 0 \\ 0 & \rho_{22} \end{array} \right) \left( \begin{array}{c} \dot{\phi_{1}}  \\ \dot{\phi_{2}} \end{array}\right)& = 
\left( \begin{array}{cc} \rho_{11}(\hat{T}+\hat{V}_{ext}) & 0 \\ 0 & \rho_{22}(\hat{T}+\hat{V}_{ext}) \end{array} \right) 
\left( \begin{array}{c} \phi_{1} \\ \phi_{2} \end{array} \right) - 
\left( \begin{array}{cc} 0 & \mu_{12} \\ \mu_{21} & 0 \end{array} \right) 
\left( \begin{array}{c} \phi_{1} \\ \phi_{2} \end{array} \right) \nonumber \\
&+ g\left(  \begin{array}{cc} \rho_{1111}|\phi_{1}|^{2} + \rho_{1221}|\phi_{2}|^{2} & \rho_{1212}\phi_{2}^{\ast}\phi_{1} \\ 
\rho_{2121}\phi_{1}^{\ast}\phi_{2} & \rho_{2112}|\phi_{1}|^{2} + \rho_{2222}|\phi_{2}|^{2} \\ \end{array} \right)
\left( \begin{array}{c} \phi_{1} \\ \phi_{2} \end{array} \right).
\end{eqnarray}

In the above, we also use the fact that, for unitary time evolution, the norm-preserving Lagrange multipliers, $\mu_{11}$ and $\mu_{22}$, are unnecessary.

%\section{Choice of basis and determining the initial state}

As the art of variational science is to find a variational space that gives the desired limiting cases automatically, we show that, by using as $\phi_{j}$ what may
be considered to be generalized Wannier functions satisfying the non-linear equations (\ref{finalEOM}), we get the correct high barrier limit
of two fragmented condensates.  We discuss this in detail in Section \ref{Basis} giving a comparison to other assays at the splitting problem.
% SHOW FIG

%\section{Landau criterion for low barrier, and motivation of $g_{12}$}
Finally, we point out an external criterion with which we can ensure that our theory is consistent with correct theory in the low barrier limit (\textit{i.e.} mean 
field theory as encompassed in the condensate order parameter and Gross-Pitaevskii equation).  For a zero-temperature Bose-Einstein condensate, all physical data 
resides in the complex order parameter $\Phi_{OP}(r,t)$ which satisfies the relationship $\Phi_{OP}(r, t) = \sqrt{\rho_{S}(r, t)}e^{i\Theta(r, t)}$, where $\rho_{S}(r, t)$ 
is the spatial density of bosons \cite{Landau}.  As such, even in the case when a compound or parametrized computational basis is used, the above relation can be used to 
check the consistency of our ansatz.

We further use the sufficiency of the complex order parameter to determine a renormalized interaction strength as a means to compensate for any errors introduced by 
the possibly ``spurious'' kinetic energy terms in the generalized Wannier basis or from using the uncoupled Fock-state equations for the $\{ \phi_{k} \}$.  The procedure 
we adopt is to use a diagonal and off-diagonal interaction strength, $g$ and $g_{12}$ respectively, in order to describe the strength of the particle-particle interactions.  
To keep consistency 
with the high-barrier limit of two independent condensates, the $|\phi_{k}|^{2}\phi_{k}$ terms use the prefactor $g$, however, we mediate the strength of the 
$|\phi_{j}|^{2}\phi_{k} \,\,(k \neq j)$ by a $g_{12}$ chosen to give the correct density, and therefore order parameter, at zero barrier.
While this procedure introduces an \textit{ad hoc} modification to the equations of motion, we note that, because of the sufficiency of the order parameter, this does 
generate the exact ground state of the system when correctly interpreted. In practice, we find that the modulation of $g_{12}$ is on the order of a few to tens of 
percent and has no significant effect on the basic scientific conclusions derived from this computational theory.  

With a final rearrangement of terms and introduction of the renormalized parameter $g_{12}$, 
Our final equations of motion, a coupled system of equations for the Fock and spatial dynamical variables, therefore, appear as:
\begin{equation}
i\hbar\dot{C}_{\gamma} = \sum_{\beta = 0}^{N}\cal{H}_{\gamma \beta}C_{\beta}
\end{equation} for the $\{ C_{\alpha} \}$, with $\cal{H}_{\gamma \beta}$ defined in (\ref{FockHam}) and 

\begin{eqnarray}\label{finalg12EOM}
i\hbar\left( \begin{array}{cc} \rho_{11} & 0 \\ 0 & \rho_{22} \end{array} \right) \left( \begin{array}{c} \dot{\phi_{1}}  \\ \dot{\phi_{2}} \end{array}\right)& = 
\left( \begin{array}{cc} \rho_{11}(\hat{T}+\hat{V}_{ext}) & 0 \\ 0 & \rho_{22}(\hat{T}+\hat{V}_{ext}) \end{array} \right) 
\left( \begin{array}{c} \phi_{1} \\ \phi_{2} \end{array} \right) - 
\left( \begin{array}{cc} 0 & \mu_{12} \\ \mu_{21} & 0 \end{array} \right) 
\left( \begin{array}{c} \phi_{1} \\ \phi_{2} \end{array} \right) \nonumber \\
&+ \left(  \begin{array}{cc} g|\phi_{1}|^{2}\rho_{1111} & 2g_{12}\phi_{2}^{\ast}\phi_{1}\rho_{1212} \\ 
2g_{12}\phi_{1}^{\ast}\phi_{2}\rho_{2121} & g|\phi_{2}|^{2}\rho_{2222} \\ \end{array} \right)
\left( \begin{array}{c} \phi_{1} \\ \phi_{2} \end{array} \right)
\end{eqnarray}
Where in the above the renormalized parameter $g_{12}$ has been introduced and the identity $\rho_{1221} = \rho_{1212}$ has been used to allow a useful 
rearrangement of terms.

\subsection{Interpretation of State Vector}

Despite explicitly using two spatial basis functions $\{ \phi_{1}, \phi_{2} \}$ in this theory and a full breadth of Fock-space variables $\{ C_{\alpha} \}$, we are 
interested in determining whether the system is characterized as two-mode vs. one-mode.  In order to do this, we follow \cite{zoller} and use the Onsager-Penrose 
criterion.  By tracing over the spatial variables of the density operator one gets a $2 \times 2$ matrix.

\begin{equation}\label{RhoF}
\rho_{F} = \left( \begin{array}{cc}
\langle a_{1}^{\dagger}a_{1} \rangle & \langle a_{1}^{\dagger}a_{2} \rangle \\
\langle a_{2}^{\dagger}a_{1} \rangle & \langle a_{2}^{\dagger}a_{2} \rangle
\end{array} \right)
\end{equation}
This reduced density operator has two eigenvalues $\Lambda_{+/-}$ which sum to $N$.  In the case that $\Lambda_{+} = N$ and $\Lambda_{-} = 0$, the system could be 
sufficiently described by the GP equation.  Should the Fock-space density operator yield eigenvalues $\Lambda_{+} = \Lambda_{-} = N/2$, then the system is ``fragmented''
into two independent condensates.  Finally, a correlated two-mode model, such as has been used to describe Josephson junctions \cite{Smerzi}\cite{Mahmud2well} and BEC atom 
interferometers \cite{InterfLim}\cite{InterfDephase}, is only appropriate when $N > \Lambda_{+} > N/2$.

The other reduced density matrix, i.e. the density matrix traced over the Fock-space variables, returns the spatial density of the system and is given explicitly by
\begin{equation}\label{RhoS}
\rho_{S} = \langle a_{1}^{\dagger}a_{1} \rangle|\phi_{1}|^2 + 2\Re\lbrace\langle a_{1}^{\dagger}a_{2}\rangle\phi^{*}_{1}\phi_{2}\rbrace + \langle a_{2}^{\dagger}a_{2} \rangle |\phi_{2}|^2
\end{equation}
this is the quantity we use, along with the Landau's identification, which becomes:
\begin{equation}\label{landauident}
\Phi_{OP}(r, t) = \sqrt{\rho_{S}(r, t)}e^{i\Theta(r, t)}
\end{equation}
in order to verify that our two-mode basis correctly reproduces all the physical data of the mean-field state when the mean-field description of the condensate is appropriate.

% IMPLEMENTATION SECTION
\section{Implementation}\label{Implementation}
In this section we describe how we solve and implement the theory described above.  Because this paper, the first of a two-part series, will be concerned with comparing the 
stationary states of various theories, this amounts to diagonalizing the equations of motion described above for the various basis sets under consideration in Section 
\ref{Basis}.  Time evolution and applications of this formalism to barrier raising and ballistic expansion will be the subject of Part II.  We find the ground state of the system by employing a relaxation method to the complex time version of the equations discussed above.  This is done by 
taking the Wick-rotated equations of motion $t \rightarrow \tau = - i t$ and adding self-consistent estimates of the chemical potential and system energies from the 
spatial and Fock equations of motion respectively.  The Wick-rotated equation of motion for either variable are:

\begin{eqnarray}\label{ComplexTimeS}
-\hbar\left( \begin{array}{cc} \rho_{11} & 0 \\ 0 & \rho_{22} \end{array} \right) \left( \begin{array}{c} \frac{\partial \phi_{1}}{\partial \tau}  \\ 
\frac{\partial \phi_{2}}{\partial \tau} \end{array}\right)& = 
\left( \begin{array}{cc} \rho_{11}(\hat{T}+\hat{V}_{ext}) & 0 \\ 0 & \rho_{22}(\hat{T}+\hat{V}_{ext}) \end{array} \right) 
\left( \begin{array}{c} \phi_{1} \\ \phi_{2} \end{array} \right) - 
\left( \begin{array}{cc} \mu_{11} & \mu_{12} \\ \mu_{21} & \mu_{22} \end{array} \right) 
\left( \begin{array}{c} \phi_{1} \\ \phi_{2} \end{array} \right) \nonumber \\
&+ \left(  \begin{array}{cc} g|\phi_{1}|^{2}\rho_{1111} & 2g_{12}\phi_{2}^{\ast}\phi_{1}\rho_{1212} \\ 
2g_{12}\phi_{1}^{\ast}\phi_{2}\rho_{2121} & g|\phi_{2}|^{2}\rho_{2222} \\ \end{array} \right)
\left( \begin{array}{c} \phi_{1} \\ \phi_{2} \end{array} \right)
\end{eqnarray}

\begin{equation}\label{ComplexTimeF}
\hbar\frac{\partial C_{\gamma}}{\partial \tau} = E_{0} - \sum_{\beta = 0}^{N}\cal{H}_{\gamma \beta}C_{\beta}
\end{equation}

%Denoting the system for the spatial variables as $\frac{\partial\phi_{j}}{\partial \tau} = {\cal D}\{ \}$
Now, instead of the evolution characteristic of the Schr\"{o}dinger equation $\phi_{j}(t) \propto e^{-it\mu/\hbar}$ and $C_{\alpha}(t) \propto e^{-itE/\hbar}$,
the equations (\ref{ComplexTimeS})(\ref{ComplexTimeF}) evolve as a form of exponential decay when the state has higher energy/chemical potential than the true ground 
state:
$\phi_{j}(\tau) \propto e^{\tau(\mu_{jj}-\mu)/\hbar}$ and $C_{\alpha}(\tau) \propto e^{\tau(E_{0}-E)/\hbar}$
So that, as the quantities $\mu_{jj}$ and $E_{0}$ iteratively converge to the ground state of the chemical potential and system energy, respectively.  This is done by 
updating these quantities by the identifications $\mu_{jj} = i \hbar \int\!dx \,\phi_{j}^{\ast}\dot{\phi_{j}}$ and 
$E_{0} = \sum_{\alpha, \beta} C_{\alpha}^{\ast} {\cal H}_{\alpha, \beta} C_{\beta}$, for the provisional values of $\phi$ and $C$, renormalized after a few e-foldings
of the relaxation process. 

When considering the excited states of the \textit{gerade}, \textit{ungerade} basis discussed in Section \ref{Basis}, the configuration of $C_{\alpha}$
for each such excited state is known \textit{a priori}, and in this case, for a fixed distribution of $C_{\alpha}$, the system (\ref{ComplexTimeS}) will unambiguously
converge to the functions and chemical potential for that excited state and not the true ground state of the system.

One technical point is that since the system of equations comprised of (\ref{ComplexTimeS}), and (\ref{ComplexTimeF}) is non-linear, there is no guarantee that a 
stationary state produced by this method is a true ground state of the system.  Stated another way, non-linearity means that there is no single energy surface for these 
variables since the energy surface depends on the $\{\phi_{j}\}$ themselves.  In order to ensure that we have not converged to such a ``non-linear local minimum,'' we 
perform this relaxation method on several different guesses for the initial conditions and verify that they converge to the same answer.

% END REWRITTEN THEORY SEC
\section{Basis Comparison}\label{Basis}

\subsection{Numerics}
The typical choice in discussing the splitting problem is to select basis functions of \textit{gerade} and \textit{ungerade} symmetries and use them to create approximately 
left and right-localized functions in order to describe the high barrier state.  This was first done in the context of the statics of a weakly-interacting gas 
in a double well potential in \cite{SandS} using the ground ($\phi_{g}$) and first excited state ($\phi_{u}$) solutions of the Schr\"{o}dinger equation and their 
linear combinations $\phi_{L} = \frac{1}{2}(\phi_{g} + \phi_{u})$ and $\phi_{R} = \frac{1}{2}(\phi_{g} - \phi_{u})$.  This has the appeal that, in the non-interacting, 
static limit, at zero barrier $(\phi_{L} + \phi_{R})^{N} = \phi_{g}^{N}$, the true many-body ground state of the system.

More recently, the dynamics of the splitting process has been investigated using the $\lbrace \phi_{g}, \phi_{u} \rbrace$ basis directly with the physical identification
that if both \textit{gerade} and \textit{ungerade} functions are macroscopically occupied, the system is ``fragmented'' \cite{ced_well}\cite{SchmiedGrond}.

We use neither of these methods and, in this section, point out some pathologies associated with either the \textit{gerade}, \textit{ungerade} scheme.  
Instead we show that a basis which can be considered to be the non-linear generalization of Wannier basis functions generates a theory with a level crossing as
the central barrier increases and show that this is a generic feature of that basis.

We show that no such level crossing exists in the $\lbrace \phi_{g}, \phi_{u} \rbrace$ basis for the same system and give a criterion for when the \textit{ungerade} states 
become important as the system trends towards the thermodynamic limit.
 
Considering a system of $200 ^{87}Rb$ atoms confined in a trap with frequencies $\omega_{x} = 2\pi \times 44.7 Hz$, 
$\omega_{y} = \omega_{z} = \omega_{\perp} = 2\pi \times 1.1 kHz$ in the presence of a central Gaussian barrier characterized by a variance of $10 \rm{\mu m}$, we compute the 
energies of various configurations of interest.  The complex time versions of the effective $1D$ equations of motion developed above in sections \ref{Implementation} 
and \ref{effective1d} are implemented in order to perform this study.

\begin{figure}
\includegraphics[scale=0.25, angle=-90]{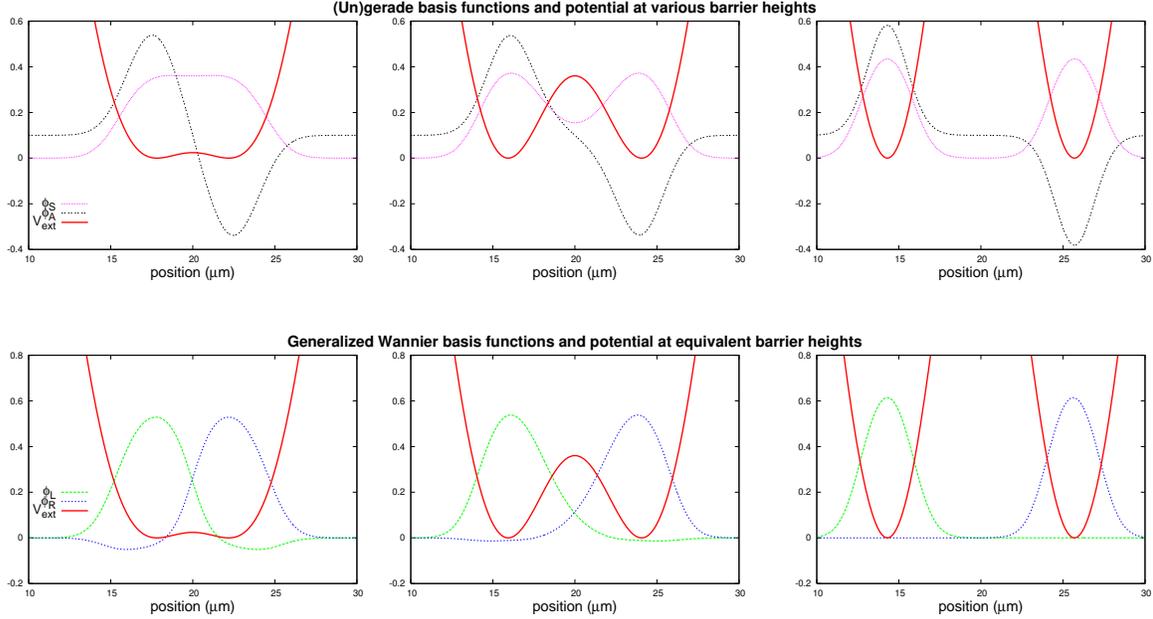}
\caption{\label{basispic}Picture of basis functions used in \cite{ced_well} and this work}
\end{figure}

\begin{figure}
\includegraphics[scale=0.25, angle=-90]{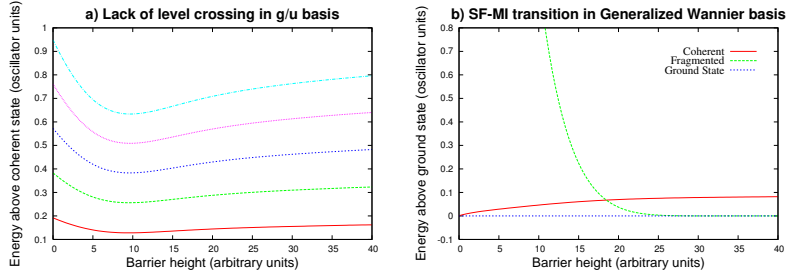}%
\caption{\label{engyfig}(Non-)existence of curve crossings in de-localized and Wannier basis theories.  In a) the heights of various excited states 
(as defined by promoting an additional particle from $g$ to $u$) above the ground state (all particles in $g$) is shown.  No crossing is observed in this 
basis, indicating that observed population of low-barrier ``excited'' states in $g$/$u$ models may be a function of equations of motion which require initial 
population of the ground state.  In b), we observe 
an energy level crossing consistent with the SF-MI transition in the Wannier basis.  In this panel, the two-mode coherent state is defined by a Binomial distribution 
of $|C_{\alpha}|^{2}$ and self-consistently relaxed $\phi_{j}$, and the ``fragmented'' state is a single, unit $C_{\alpha = N/2}$ as described in the text.}
\end{figure}

\subsection{Analytic model of energy crossings}
We present a discussion, in the context of this model, of generic features of either choice of basis and show that the Wannier basis always has a Mott Insulator
ground state at high barriers while the \textit{gerade}, \textit{ungerade} does not.  Further, it can be shown that the \textit{ungerade} state becomes irrelevant
in the large-$N$ limit, as the energy gap scales as $gN$ - an alternate proof of the perturbative result of Huang and Yang \cite{HuangYang}.

In order to show that the Mott Insulator state is a generic feature of the Wannier basis, we show that the Hamiltonian expressed in the Fock basis (\ref{FockHam}) 
has a minimum at $N_{1} = N/2$ and is concave.  In the high-barrier limit, as in lower-right panel of Fig \ref{basispic}, when the basis functions are localized 
$\displaystyle \int\!dr\, |\phi_{1}|^{2}|\phi_{2}|^{2} \rightarrow 0$, the (otherwise pentadiagonal) Fock-basis Hamiltonian, becomes diagonal:
\begin{equation}
{\cal H}_{\beta \beta} \equiv \langle \beta | \left\{ \sum_{j=1}^{2} \epsilon_{jj}(\hat{a}_{j}^{\dagger}\hat{a}_{j}) + 
\frac{g}{2}\sum_{j=1}^{2}\Gamma_{jjjj}(\hat{a}_{j}^{\dagger}\hat{a}_{j}^{\dagger}\hat{a}_{j}\hat{a}_{j}) \right\} | \beta \rangle
\end{equation}
and the energy becomes
\begin{equation}
\langle {\cal H} \rangle = \sum_{\beta=0}^{N}C_{\beta}^{\ast}{\cal H}C_{\beta} = \\
|C_{\beta}|^{2}(\epsilon_{11}\beta + \epsilon_{22}(N-\beta)+ \frac{g}{2}\Gamma_{1111}\beta(\beta - 1) +\frac{g}{2}\Gamma_{2222}(N-\beta)(N - \beta + 1))
\end{equation}
Invoking the symmetry of the well, $\Gamma_{1111} = \Gamma_{2222}$ for instance, the energy then can be written 
\begin{equation}
\langle {\cal H} \rangle = \sum_{\beta=0}^{N}C_{\beta}^{\ast}{\cal H}C_{\beta} = 
|C_{\beta}|^{2}(\epsilon_{11}N + \frac{g}{2}\Gamma_{1111}(2\beta^{2}-2N\beta+N^{2}-N))
\end{equation}
Which is concave and has a minimum with respect to $\beta$ at $\beta = N/2$.  Finally, since ${\cal H}$ is diagonal when the functions $\phi_{j}$
don't overlap, the ground state is a single configuration $C_{\beta} = \delta_{\beta, N/2}$, or Fock state and the system is in a Mott Insulator phase.

Conversely, there is no analogue for this proof for the de-localized functions $\{ \phi_{g}, \phi_{u} \}$, or their nonlinear generalizations.  We can, however,
examine how relevant the excited states, as defined by the promotion of an additional particle from $\phi_{g}$ to $\phi_{u}$ are in the large $N$ limit.  Looking 
at the energy difference between all particles in $\phi_{g}$ (since the quantities in (\ref{FockHam}) depend on particle number, we denote the ground state
factors of $\epsilon$ and $\Gamma$ by $\circ$) and $N-1$ particles in $\phi_{g}$ one particle in the first excited state (quantities denoted by $+$):
\begin{eqnarray*}
\lefteqn{\Delta E = \langle {\cal H} \rangle_{+} - \langle {\cal H} \rangle_{\circ} = 
(\epsilon_{uu}^{+} + (N-1)\epsilon_{gg}^{+})} \\ 
& & + \frac{g}{2}((N-1)(4\Gamma_{gugu}^{+}) + (N-1)(N-2)\Gamma_{gggg}^{+}) - (N\epsilon_{gg}^{\circ} + \frac{g}{2}N(N-1)\Gamma_{gggg}^{\circ})
\end{eqnarray*}
Since we are simply looking for a scaling argument we can equate the geometric quantities which only differ by the exclusion of a single particle
(for instance $\Gamma_{gggg}^{+} = \Gamma_{gggg}^{\circ}$) in the limit of $N \gg 1$ giving.
\begin{equation}
\Delta E = \epsilon_{uu} - \epsilon_{gg} + \frac{g}{2}(4(N-1)\Gamma_{gugu} + (2-2N)\Gamma_{gggg})
\end{equation}
Roughly, one can estimate this by noting that $\Gamma_{gugu}$ and $\Gamma_{gggg} \sim 1$ and $\epsilon_{uu} - \epsilon_{gg} \sim \hbar\omega$
giving 
\begin{equation}
\Delta E \sim \hbar\omega + gN
\end{equation}

A complimentary derivation of the energy gap for a dilute Bose gas first published in \cite{HuangYang} 
and strong indication that the energy crossings and population of excited states reported 
in computational attempts on the splitting problem which use a de-localized basis \cite{ced_well}\cite{SchmiedGrond} may only be relevant at low particle numbers and 
those results may not generalize to large-$N$ systems.  We further point out that using these methods to compute factors of $\Gamma$ and $\epsilon$ exact spectra for 
arbitrary trapping potentials and particle numbers can be calculated when the de-localized basis is used.  

We make two remarks here about the presence of this energy gap in models which employ a de-localized basis.  First, this calculation shows that the alternate definition of 
``fragmentation'' which has appeared in the literature, namely, that a condensate is ``fragmented'' when both the $g$ and $u$ states are macroscopically occupied 
is not equivalent to the definition of a single unit-probability partition of the particles in between wells which is native to Bose-Hubbard type models.  The former
definition corresponds to a state which is, generically, of more energy than the ground state of that theory (all particles in $g$) and therefore must be reached by a 
non-adiabatic process.  Conversely, we have shown that the Mott insulator ground state is a generic feature when Wannier basis methods are used to model repulsive 
condensates.

Secondly, in the context of dynamical investigations of the splitting process, the non-invertability of $\rho_{jk}$ (as discussed in Sec. \ref{Theory}) means that in 
order for the equations of motion such as those used in \cite{CedTD} and \cite{HDMreview} to be nonsingular, the initial state must include some finite population of 
an energy level above this gap which is increasingly unphysical in the large-$N$ limit.  We note that if the structure of the energy levels seen in Fig. \ref{engyfig} a) 
is general and the energy gaps maintain approximately-equal spacing as the barrier raises, this may explain the oscillatory behavior observed in \cite{ced_well}, for
instance.

\begin{figure}
\includegraphics[scale=0.5, angle=-90]{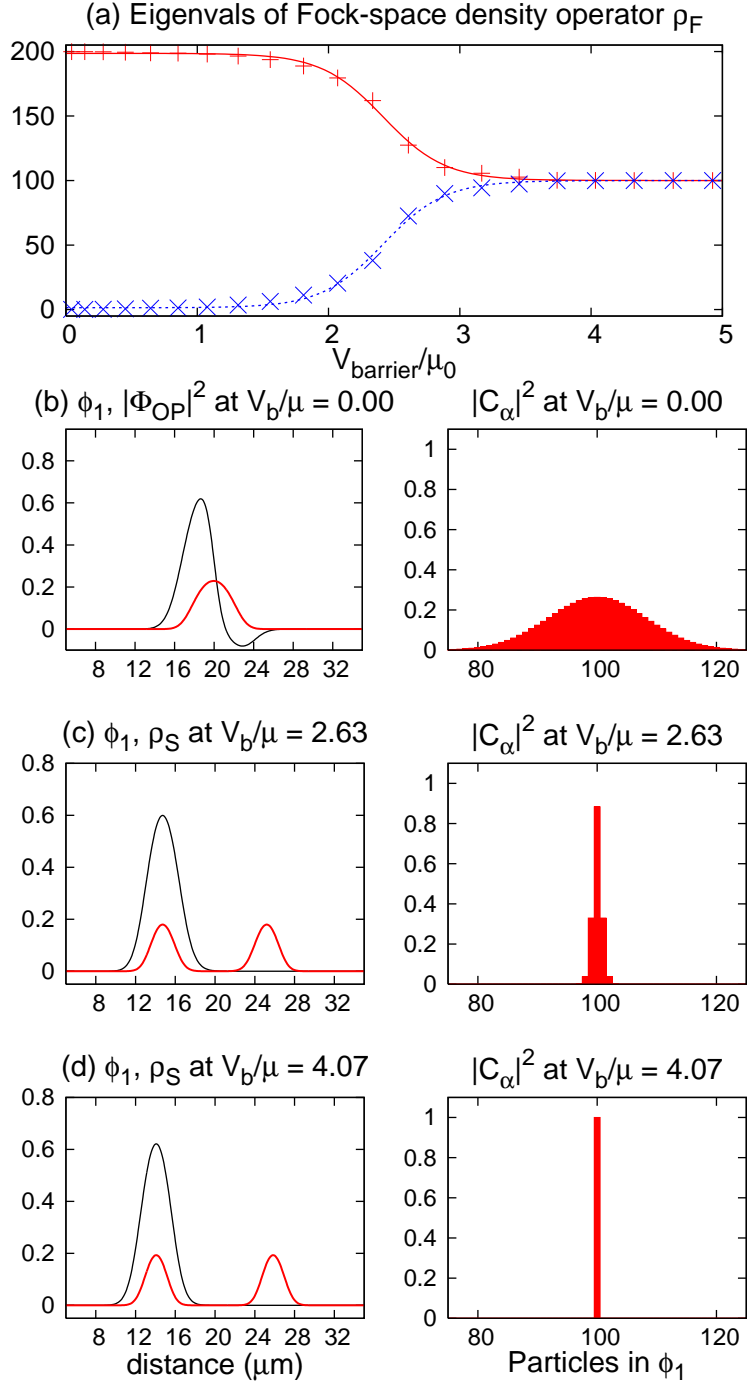}
\caption{\label{densCIplot} A selection of the physical data which comprise our theory.  In a) the system eigenvalues are shown as a function of barrier height, clearly
spanning three distinct regimes.  In b) the computational variables from zero-barrier state is shown, only the final density has direct physical meaning.  In c) the data 
from the two-mode regime is shown, both the $\phi_{j}(r)$ and $C_{\alpha}$ have the interpretation of the spatial and Fock-space variables of a two-mode theory.  In d) 
the data for a ``fragmented'' state is shown in which both $\phi_{1}$ and $\phi_{2}$ have the interpretation of independent complex order parameters and there is no 
relative particle number uncertainty.}
\end{figure}

As a graphical summary of the physical interpretation we propose for our theoretical description of the splitting process, we include figure \ref{densCIplot}.  In 
this figure, we again find the ground state of the theory outlined in \ref{Theory} and \ref{Implementation} for the $200 ^{87}Rb$ atom system described above.  This figure
shows relevant physical data for a range of barrier heights.  Clearly visible in Fig \ref{densCIplot}a) are three distinct regimes, a coherent regime in which all $N$
particles reside in a single state and GP theory should be applicable; a two-mode state, where there are two populated eigenvalues and it is sensible to talk about 
relative populations of two mode functions; and finally a regime in which there are two ``fragmented'' BECs which are described by who independent complex order 
parameters.

As discussed in the figure caption, the eigenvalues show how to interpret the state.  When there are $\lbrace N,0 \rbrace$ eigenvalues, only the final density 
(and phase, should one be doing dynamics) has a clear physical interpretation through the Landau identification (\ref{landauident}) and the 
$C_{\alpha}$ and $\phi_{j}$ are simply elements of a computational basis.  On the shoulder of the best-fit \textit{tanh} functions, a two-mode model is 
appropriate and the $\phi_{j}$ become the mode functions.  In this case, the $C_{\alpha}$ also have a direct physical interpretation as $|C_{\alpha}|^{2}$ being
the probability of detecting $\alpha$ particles in $\phi_{1}$.  Finally, when the system has two $N/2$ eigenvalues, only one $C_{\alpha}$ has non-zero probability
and the system is correctly described as two independent condensates.  If this were at the end of a dynamical splitting process, one would say that the initial 
condensate has been ``fragmented'' into two condensates.

We have shown that the basis we have selected to generate the theory of a splitting BEC correctly displays a curve crossing in energies corresponding to the 
Superfluid to Mott insulator transition and that this is a generic feature of the model.  We have also shown three distinct physical regimes of the splitting BEC
and have carefully delineated what parts of the computational basis have a clear physical interpretation in each of these regimes.

\section{Analytics and Numerics of the Relevance of a Two-Mode Model}\label{Results}

We use the formalism developed in section \ref{Theory} to determine the ground state of a variable number of bosons in the trap discussed above.  Our goal is 
to understand how various aspects of the splitting process scale with the total particle number $N$ and what generic features can be observed.  
We observe that a type of universal behavior is associated with the splitting of a dilute Bose-condensed gas and we present a simple analytic model to understand 
this behavior.  Finally, we discuss the ramifications of this universal feature of a splitting BEC.

\subsection{Numerics}
As in the preceding section, we use the complex-time form of the equations shown above and a relaxation algorithm in order to compute the ground state of a system of 
various values of $N$ $^{87}Rb$ atoms confined in a external potential characterized by trap frequencies $\omega_{x} = 2\pi \times 44.7 Hz$, 
$\omega_{y} = \omega_{z} = \omega_{\perp} = 2\pi \times 1.1 kHz$ and a central Gaussian barrier, applied along the $x$ direction (See \ref{effective1d} for details
of how this system was modeled as an effective 1D system).

\begin{figure}  
\includegraphics[scale=0.35, angle=-90]{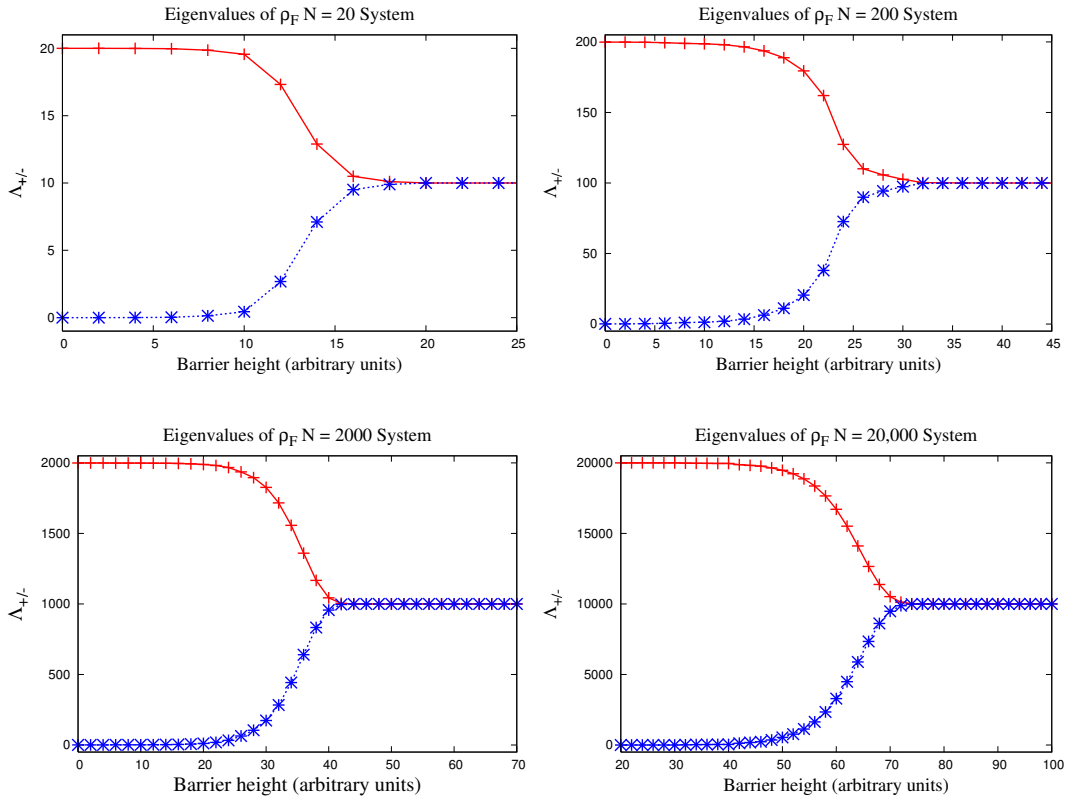}
\caption{\label{EntFig} Eigenvalues of the Fock-space density operator as a function of barrier height for four decades of $N$. In each case, each of the three
regimes discussed in the text corresponds to a finite range of barrier heights.}
\end{figure}

In accordance with discussion in \ref{Theory}, in order to discern in between a single-mode ``GP-like'' state and a system that truly requires a two-mode description, 
the quantity of interest is the density operator traced over the spatial degrees of freedom or the ``Fock-space density operator.'' (\ref{RhoF}).  We plot the eigenvalues 
of (\ref{RhoF}) as a function of increasing barrier height for four decades of particles, $N =  \{20,\, 200,\, 2000,\, 20000\}$, in the trapping potential described above.  
The results of these calculations are shown in Fig \ref{EntFig}.

In the context of the splitting of a BEC as a phase transition, this figure may be understood as the left most region of each graph, characterized by eigenvalues of
$\Lambda_{+/-} = \lbrace N, 0 \rbrace$ as being a ``GP-like'' state, namely, the Gross-Pitaevskii equation is sufficient to describe 
the zero-temperature physics of this state.  
At high barriers, when the state is characterized by $\Lambda_{+/-} = N/2, N/2$, the two generalized Wannier functions are separated enough that an arbitrary phase
shift on one of them, will not affect the total system energy, satisfying the requirement that this process be a type of phase transition.  The region in between these
two asymptotes, or phases, is where it is appropriate to use a two-mode model to discuss the relevant physics.

Since the production of the first dilute alkali gas BEC, schemes have been proposed in which atom interferometry is performed on an initially phase-coherent two-mode BEC, 
see for instance \cite{MeystreUys} \cite{BurnettBECinterf}, therefore it is of fundamental interest to understand when a two-mode model is relevant.  We devote the rest of 
this paper to understanding under what circumstances a single coherent, BEC can be brought into the two-mode regime.  

First, in order to examine the splitting process in a system-independent manner we define the 
intensive quantity we call the \textit{entanglement} of the system which is simply equal to the normalized difference in between the eigenvalues of (\ref{RhoF}): 
\textit{entanglement} $= (\Lambda_{+} - \Lambda_{-})/N$.

\begin{figure}  
\includegraphics[scale=0.4, angle=-90]{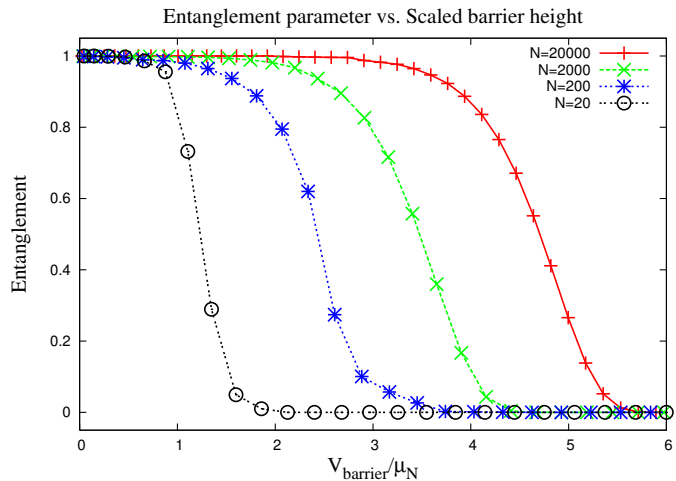}
\caption{\label{ScaleEntFig} Scaled graphs of the \textit{entanglement} which suggest a lack of non-analyticity in the splitting process as $N \rightarrow \infty$.}
\end{figure}

In Figure \ref{ScaleEntFig} we show the \textit{entanglement} plotted versus the height of the central barrier for a few decades of total particle number, showing the 
transition from a unit-\textit{entanglement} ``GP-like'' state through a system correctly described by a two-mode model and finally to two independent condensates described
by \textit{entanglement} zero.  We further roughly scale out the dependence of the barrier height for different particle numbers by plotting the \textit{entanglement} versus 
$V_{barrier}(x=L/2)/\mu_{N}$, where $\mu_{N}$ is the chemical potential of each system, $N = \{20,\, 200,\, 2000,\, 20000\}$, at zero barrier.  

While it is typical to expect some non-analyticity to emerge in the thermodynamic limit, Fig \ref{ScaleEntFig} suggests, however a universal feature associated with 
the process of transitioning from a one-mode state through a two-mode state in order to fragment a BEC.  We conclude that this is, in fact, the case and present an 
analytic model to supplement this numeric data in the next subsection.

\begin{figure}  
\includegraphics[scale=0.5, angle=-90]{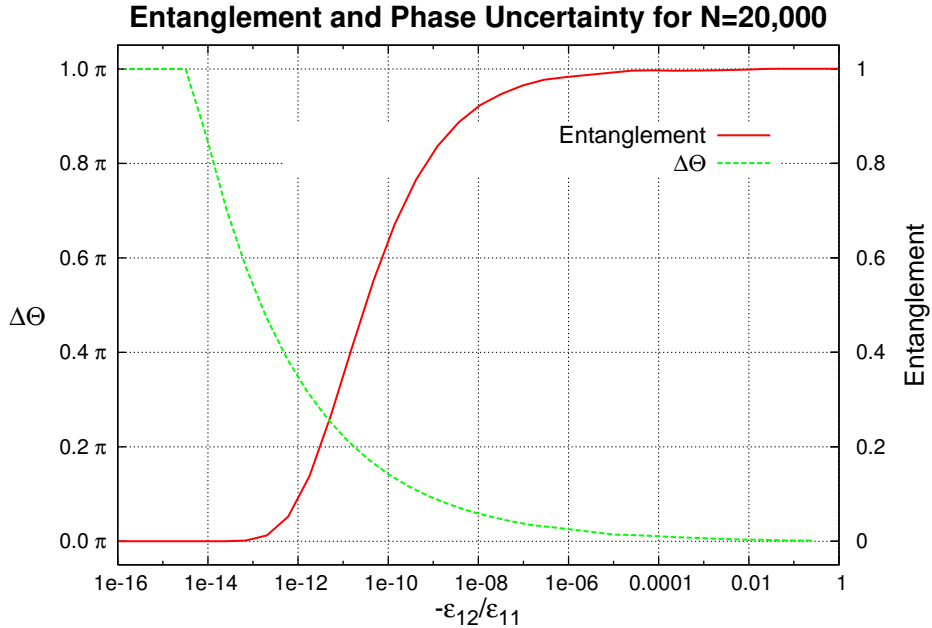}
\caption{\label{HopEntFig} Logarithmic scale graph of the \textit{entanglement} as a function of (unitless) tunneling rate for $N = 20,000$.}
\end{figure}

Since the peak barrier height, even when scaled to the chemical potential, is not independent of the system geometry, we show how the \textit{entanglement} of the 
$N = 20,000$ particle system depends on the ratio of the self-energy to the hopping energy $\epsilon_{12} / \epsilon_{11}$.  These quantities, typically denoted 
$J$ and $U$, are analogous to the on-site and hopping strengths of lattice Bose-Hubbard type models\cite{FisherSFMI}.  Figure \ref{HopEntFig} shows that, while a 
two-mode model is appropriate over several decades of the hopping rate before fragmentation, a single-mode GP-like state is a necessary and sufficient description 
until the ratio of on-site to hopping strength, ($\epsilon_{12} / \epsilon_{11}$) is well below $10^{-6}$.  In addition to the \textit{entanglement}, we plot the expected 
phase uncertainty of a two-mode model using the Heisenberg relation $\Delta N \Delta \Theta \approx 1$ \cite{AndersonRMP} as a rough guide to when two-mode 
interferometry would fail for lack of a well-defined relative phase.

\subsection{Analytic model of splitting regime}
In order to understand this behavior, we present an analysis of the basic mathematical quantity which dictates splitting process 
by assuming a simple analytical form for the distribution of the $\{ C_{\alpha} \}$.  Examining
the Fock-space density operator (\ref{RhoF}), for a symmetric geometry, the diagonal entries are known:  
$\langle a_{1}^{\dagger}a_{1} \rangle = \langle a_{2}^{\dagger}a_{2} \rangle = N/2$.  Therefore the \textit{entanglement} is determined solely by the off-diagonal 
entry $\rho_{12} = \langle a_{1}^{\dagger}a_{2} \rangle$ and its complex conjugate.  Explicitly, this is quantity is given by 
\begin{equation}\label{Faustian}
\rho_{12} = \displaystyle \sum_{\alpha, \beta = 0}^{N} \langle \alpha | C_{\alpha}^{\ast} (a_{1}^{\dagger}a_{2})  C_{\beta} | \beta \rangle
 = \displaystyle \sum_{\alpha = 1}^{N} C_{\alpha}^{\ast}C_{\alpha-1}\sqrt{N-\alpha+1}\sqrt{\alpha}
\end{equation}
When this quantity vanishes, the state is fragmented or two independent condensates and when it equals $N/2$ the system is at unit \textit{entanglement}, and a description
by the GP equation is appropriate.  The sum in (\ref{Faustian}) is a weighted, unit-offset autocorrelation of the distribution $\{ C_{\alpha} \}$ and, by assuming an 
physically-motivated analytic form of $\{ C_{\alpha} \}$, we are able to get an insight into the process of fragmenting a BEC.

Assuming a continuous, Gaussian probability distribution of $|C_{\alpha}|^{2} = \frac{1}{\sqrt{2 \pi \sigma^{2}}} e^{-\frac{(\alpha - N/2)^{2}}{2 \sigma^{2}}}$, the limits 
discussed above are represented by means on the variance of the distribution, i.e. $\sigma \approx \sqrt{N}$ for low-barrier ground states, well-described by GP theory
and $\sigma \approx 1$ in the fragmented limit.  

\begin{figure}
\includegraphics[scale=1.0, angle=-0]{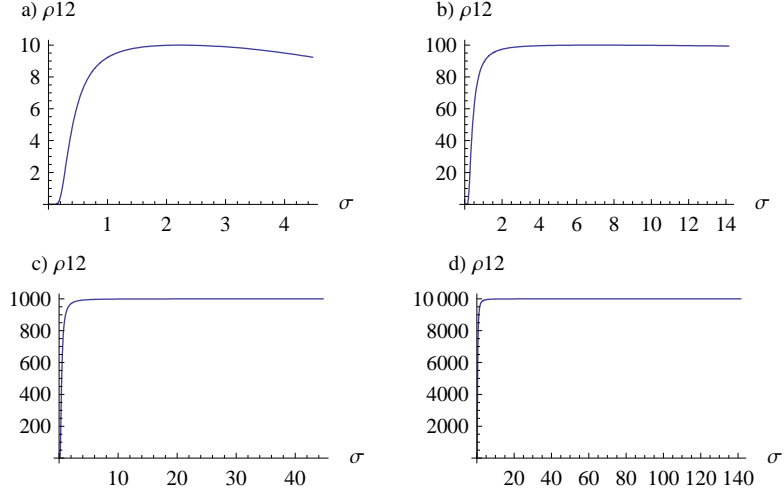}
\caption{\label{EntGrid}Graph of $\rho_{12}(2/N)$ vs. width of the Gaussian ansatz $C_{\alpha}$ distribution spanning four decades of $N$ showing the regimes of 
\textit{entanglement} a) $N = 20 \; ^{87}Rb$ atoms, b) $N = 200 \; ^{87}Rb$ atoms, c) $N = 2000 \; ^{87}Rb$ atoms, d) $N = 20,000 \; ^{87}Rb$ atoms.}
\end{figure}

In Fig \ref{EntGrid}, we show the how the value of $\langle a_{1}^{\dagger}a_{2} \rangle$ depends on $\sigma$ for a few decades of total particle number $N$.  
In Fig \ref{EntGrid}a), the failure of $\rho_{12}$ to stay at the correct asymptotic value of $N/2$ is due to the tails of the Gaussians exceeding the physical boundaries 
of the integration, $\alpha \in \{0, N\}$ and is an artifact of using this approximation for small particles numbers.

\begin{figure}
\includegraphics[scale=0.7, angle=-0]{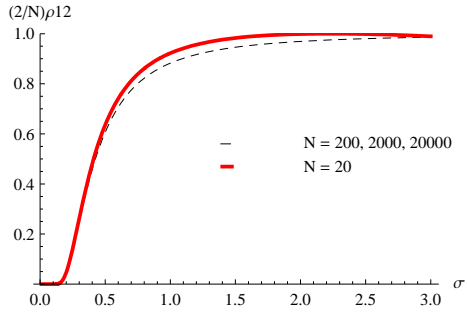}
\caption{\label{Universal}Universal feature of $(2/N)\rho_{12} =$ \textit{entanglement} as desribed by the Gaussian model discussed in the text.}
\end{figure}

Each of the graphs shows similar behavior as the system goes towards the fragmented limit.  In order to see this in Fig \ref{Universal} all four of the preceding graphs of
$\rho_{12}$ has been scaled to unity and shown in the low-$\sigma$ limit.

Since a two-mode model is only an appropriate description of a double-well condensate when the \textit{entanglement} is less than unity, this analysis indicates that 
a large particle number BEC, starting in ground state of a harmonic trap, must be very precisely controlled in order to be a true two-mode system.  Specifically,
$\sigma \approx 2$ has the physical meaning of a system which is well described by approximate Fock-states with an uncertainty in particle number of only a 
few particles since $2 \sigma \approx \Delta N$.  Condensates in splitting experiments performed to date, however, are typically on the order of $\sim 10^{5}$ 
particles\cite{MITinterf}\cite{MITphasefluc}\cite{MITsqueeze}\cite{Schmied}, making it unlikely that such systems are in a regime where true two-mode physics 
is applicable for any significant amount of time.  

We have identified a universal feature of the splitting process for identical bosons, based solely on the assumption of a unimodal distribution of particles around 
a mean value, which indicates that independent of the total particle number or trap geometry, two-mode physics is only applicable when the uncertainty, 
$\Delta N$ is on the order of a single particle.  One of the primary implications of the $N$-independence of this feature is that reliably scaling up 
experiments such as the lattice SF-MI experiments \cite{SFMI} to much larger than unit filling of each well would require extremely fine control over the splitting
potential.  Further, figure \ref{HopEntFig} shows for an specific large-$N$ system that the tunneling rate corresponding to two-mode physics is very small. 

\section{Conclusions}

We have presented a tractable method to describe the splitting of a large number of indistinguishable bosons in an external trap based on using localized non-linear
functions and a novel mapping criterion to the low-barrier state instead of the typical approach of using linear combinations of non-localized functions.
This method captures all of the features of the splitting of a BEC as a phase transition from a one-mode theory whose dynamical law is the GP equation, through a 
correlated two-mode theory and finally to a fragmented condensate defined by two independent complex order parameters, giving that system a  $U(1) \oplus U(1)$
symmetry.
By exploring the region which connects the one-mode and fragmented limits numerically and with an analytic model which captures the relevant behavior of the 
splitting process in Fock space, we are able to conclude that the region in which a true correlated two-mode model is applicable corresponds to non-vanishing 
fluctuations around a well-defined number state of no more than one or two particles  - more than this
and full coherence is established between particles in the two wells and a single, de-localized, order parameter $\Phi_{OP}$, whose time-evolution is dictated by 
the Gross-Pitaevskii equation, becomes a sufficient description of the state. 

We would like to thank David J. Masiello and Kaspar Sakmann for helpful discussions about multi-configurational methods and Nathan Kutz for sharing his expertise
on numerical methods.  We are also indebted to an anonymous referee of another draft for providing a set of detailed criteria for the clear presentation of a 
variationally-derived theory.  We gratefully acknowledge funding from NSF grant PHY 07-03278.

\appendix
\section{Effective parameters for 3D system}\label{effective1d}
We discuss a method to self-consistently include trap and mean-field physics from the condensate's extent in the two transverse directions when we are 
interested in the spatial profile (and ultimately dynamics) along only one principal splitting axis.  This Appendix deals with producing effective lower-dimension
equations of motion for NLSE/GP like equations and multi-configurational systems of equations and is not concerned with systems in which the transverse 
confinement is on the order of the healing length or scattering length, at which point it has been shown \cite{olshanii} that a modified form of s-wave scattering
takes place and an effective $a_{s}$ needs to be used.  One can verify that none of the systems under consideration in this paper are in this limit, however,
correct accounting of transverse degrees of freedom have an important effect on the results given above.

For the sake of simplicity, we present this analysis by creating an effective $1D$ equation for the $x$-direction from the full $3D$ GP equation, however it is 
exactly analogous to the treatment of the coupled mode equations above (\ref{finalg12EOM}), which have all the same essential mathematical features of the GP equation.  
In three dimensions, the GP equation may be written:
\begin{equation}
\mu\psi(x, y, z) = \left(V_{ext}(x, y, z) - \frac{\hbar^{2}}{2M}\nabla^{2} + g(N-1)|\psi(x,y,z)|^{2}\right)\psi(x,y,z)
\end{equation} 

We start by assuming a separable solution of the GP equation 
$\psi(x,y,z) = \phi(x)\psi_{\perp}(y,z)$ with each factor independently normalized $\int dx|\phi(x)|^{2} = \int dydz |\psi_{\perp}(y,z)|^{2} = 1$, the time-independent

In the commonly-discussed experiment where a barrier is raised along the $x$-direction and the transverse directions are harmonically trapped, this becomes, under our 
assumption of separability:
\begin{equation}\label{sepGP}
E\phi(x)\psi_{\perp}(y, z) = \left(V_{ext}(x) + \frac{1}{2}\omega_{y}^{2}y^{2} + \frac{1}{2}\omega_{z}^{2}z^{2} - \frac{\hbar^{2}}{2M}\nabla^{2} + g(N-1)|\phi(x)|^{2}|\psi_{\perp}(y,z)|^{2}\right)\phi(x)\psi_{\perp}(y,z)
\end{equation}
Clearly, even under the assumption of a separable \textit{solution}, the non-linear term does not allow the above equation to reduce to a set of two or three 
separable \textit{differential equations}.
  
While full $3D$ stationary and time-dependent solutions of the GP equation are regularly found\cite{soliton}\cite{BECcoherence}, we want an effective $1D$ 
equation with an eye 
towards performing tractable dynamical calculations at large particle numbers.  The solution we propose is to assume a form for the transverse solution 
$\psi_{\perp}(y,z)$ which minimizes the energy of an appropriate energy functional given an initial solution of $\phi(x)$, we then use a single value $\overline{\psi_{\perp}}$, 
such as the peak density, in the non-linear term of the - now separable - complex-time version of the GP equation which generates solutions of $\phi(x)$. 
The new solution of $\phi(x)$ will give an updated energy functional for $\psi_{\perp}(y,z)$ and the system of equations, eventually including the coupled mode and Fock-space
equations for the two-mode theory presented in the body of this paper, can be iterated to convergence.

Taking the Gaussian ansatz for the transverse profile
\begin{equation}
\psi_{\perp}(y,z) = \frac{1}{\sqrt{\pi\sigma_{y}\sigma_{z}}}e^{\frac{-y^{2}}{2\sigma_{y}^{2}}}e^{\frac{-z^{2}}{2\sigma_{z}^{2}}}
\end{equation}
We left-multiply (\ref{sepGP}) by $\psi(x,y,z)$ and integrate over all space, utilizing the fact that the $x$ and transverse factors of $\psi$ are independently normalized 
in order to construct the full energy functional
\begin{equation}
E = \epsilon^{x} + \epsilon^{\perp} + g(N-1)\Gamma^{x}\Gamma^{\perp}
\end{equation} 
\begin{center}
with the definitions $\epsilon^{x} = \int\! dx \phi^{\ast}(x)(V_{ext}(x) - \frac{\hbar^{2}}{2M}\frac{\partial}{\partial x})\phi(x)$ \\
 $\epsilon^{\perp} = \int\! dx \phi(y,z)(\frac{1}{2}\omega_{y}^{2}y^{2} + \frac{1}{2}\omega_{z}^{2}z^{2} - \frac{\hbar^{2}}{2M}(\frac{\partial}{\partial y} + \frac{\partial}{\partial z}))\psi_{\perp}(y,z)$ \\
and $\Gamma^{x} = \int\! dx \phi^{\ast}(x)\phi^{\ast}(x)\phi(x)\phi(x)$,\\ $\Gamma^{\perp} = \int dydz \psi^{\ast}(y,z)\psi^{\ast}(y,z)\phi(y,z)\phi(y,z)$
\end{center}

The ansatz must minimize the energy $E^{\perp} = \epsilon^{\perp} + g(N-1)\Gamma^{x}\Gamma^{\perp}$ or, explicitly 
\begin{equation}
E^{\perp} = \frac{\hbar^{2}}{2M}\left(\frac{1}{2\sigma_{y}^2} + \frac{1}{2\sigma_{z}^2}\right) + \frac{1}{4}(\omega_{y}^{2}\sigma_{y}^{2} + \omega_{z}^{2}\sigma_{z}^{2}) + g(N-1)\frac{\Gamma^{x}}{2\pi\sigma_{y}\sigma_{z}}
\end{equation}

Where $\phi(x)$ satisfies the effective $1D$ equation
\begin{equation}
\mu\phi(x) = \left(V_{ext}(x) -  \frac{\hbar^{2}}{2M}\frac{\partial}{\partial x} + g(N-1)\overline{\psi_{\perp}}^{2}|\phi(x)|^{2}\right)\phi(x)
\end{equation}

The two solutions can then be iterated to convergence through the parameters $\lbrace \overline{\psi_{\perp}}, \Gamma^{x} \rbrace$, and provided that the dynamics in the 
transverse directions are negligible, yield effective $1D$ time-dependent equations.

Analogous to the case of the GP equation discussed above, the trap and particle data from the transverse directions can be accounted for in the two-mode theory 
discussed above.
Summarizing the results from those calculations here $\epsilon_{jj} = \epsilon^{x}_{jj} + \epsilon^{\perp}$, $\epsilon_{jk} = \epsilon^{x}_{jk}$ if $j \neq k$,
$\Gamma_{jklm} = \Gamma^{x}_{jklm}\Gamma^{\perp}$.  Further, in the nonlinear terms of the mode equations (\ref{finalg12EOM}), a prefactor of $\overline{\psi_{\perp}}^{2}$ 
is included 
in both the diagonal and off-diagonal interaction strength.

This procedure may be further generalized in a straight-forward manner to the case when dynamics in $m$ dimensions are of interest, while the condensate is 
constrained to $D > m$ dimensions.

\bibliography{template}

\end{document}